# Superconducting properties and electronic structure of NaBi

S K Kushwaha[1], J W Krizan[1], Jun Xiong[2], T Klimczuk[3,4], Q D Gibson[1], Tian Liang[2], N P Ong[2] and R J Cava[1]

[1]*Department of Chemistry, Princeton University, Princeton NJ 08544, USA*

[2]*Department of Physics, Princeton University, Princeton NJ 08544, USA*

[3] *Faculty of Applied Physics and Mathematics, Gdansk University of Technology, Narutowicza 11/12, 80-233 Gdansk, Poland*

[4] *Institute of Physics, Pomeranian University, Arciszewskiego, 76-200 Slupsk, Poland*

**Abstract**

Resistivity, *dc* magnetization, and heat capacity measurements are reported for superconducting NaBi. $T_c$, the electronic contribution to the specific heat $\gamma$, the $\Delta C_p/\gamma T_c$ ratio, and the Debye temperature are found to be 2.15 K, 3.4 mJmol$^{-1}$K$^{-2}$, 0.78, and 140 K respectively. The calculated electron-phonon coupling constant ($\lambda_{ep}$ = 0.62) implies that NaBi is a moderately coupled superconductor. The upper critical field and coherence length are found to be 250 Oe and 115 nm, respectively. Electronic structure calculations show NaBi to be a good metal, in agreement with the experiments; the $p_x$ and $p_y$ orbitals of Bi dominate the electronic states at the Fermi Energy.

Alkali-Bi compounds have recently drawn the attention of the materials physics community due to their topological quantum properties. $Na_3Bi$ has been of particular interest as a Dirac Semimetal candidate [1-3]. While in past work several alkali-Bi compounds have been reported as superconductors [4-6], they were not well studied. NaBi is one of these superconductors, with a reported transition temperature of ~2.2 K [4]. The characterization of the superconductivity and electronic structure of this compound are of interest in spite of its low $T_c$ for two reasons. Firstly because it is a decomposition product of highly air-sensitive $Na_3Bi$, and as such its superconductivity interferes with measurements intending to characterize the topological properties of $Na_3Bi$ at low temperatures, and secondly because as a heavy metal superconductor its properties may be significantly influenced by spin orbit coupling.

Here we further characterize the superconductivity in NaBi, reporting the results of electrical transport, magnetization and specific heat characterization of the superconducting transition. DFT-based electronic structure calculations are also reported. The *dc* susceptibility and specific heat studies clearly indicate the presence of a bulk superconducting transition at 2.15 K. The field dependent electrical resistivity and magnetization measurements performed down to 0.4 K indicate a low zero temperature upper critical field of $H_{C2}(0)$ ~250 Oe in agreement with the early measurements [4]. The calculated electronic band-structure indicates that the $p_x$ and $p_y$ orbitals of Bi mainly contribute to the states near the Fermi energy, with the $p_z$ states gapped by spin orbit coupling; therefore Bi-based electronic states are responsible for the superconducting behavior of NaBi.

NaBi was synthesized by a method based on the Na-Bi phase equilibria [7]. 5N elemental Bi (Alfa Aesar, further purified by reduction in presence of C at 950 $^0C$) and Na (99.95%) were mixed in a 1:1 mole ratio and loaded into metallic tubes sealed at one end while in an Ar filled glovebox. The metallic tubes used were either carbon coated Cu, 3N pure electrical grade, or stainless steel alloy (304 L). After loading, the open tube end was crimped and arc-welded under a continuous flow of Ar. The ampoules were heated at the rate of 60 °C/h to 600 °C, held there for 12 hours to insure homogeneous mixing of the melt, and then quenched in ice-cold water. The ampoules were further annealed at 400 °C for 24 hrs. Samples were opened, handled, and prepared for all characterizations in an Ar-filled glovebox due to the high air sensitivity of the material. NaBi could be synthesized in good quality in both kinds of metal tubes.

To check the crystal structure and phase purity of the synthesized NaBi, specimens were studied by powder X-ray diffraction in a Bruker D8 FOCUS diffractometer with a Cu Kα radiation source and a graphite diffracted beam monochromator. The powdered sample was covered with Paratone-N oil and placed under a continuous flow of $N_2$ gas to minimize degradation during data recording. The diffraction pattern was recorded in the 2-theta angular range of 15–90 degrees with a step rate of 0.04°/s. The *dc* magnetic susceptibility and heat capacity were measured on a Quantum Design Physical Property Measurement System (PPMS).

The temperature and field dependence of the resistance was measured by a Linear Research Resistance Bridge, while the field dependent magnetization in the superconducting state was obtained by torque magnetometry. In these measurements, the sample is mounted on a gold cantilever and the magnetic moment generated by the anisotropic supercurrent yields a torque. As a result, the cantilever is slightly bent and there is a small change in the capacitance, measured by a GR1616 capacitance bridge, between the cantilever and the gold-plated substrate. The torque signal is given by

$$\tau = \frac{1}{2}\mu_0 V \Delta \chi^2 Sin2\theta$$

where $\Delta\chi = \chi_\perp - \chi_\parallel$ is the difference between the out-of plane susceptibility and the in-plane susceptibility. $\theta$ is the angle between the field direction and the normal direction of the sample plane.[8]

Electronic structure calculations were performed in the framework of density functional theory using the Wien2k code [9] with a full-potential linearized augmented plane-wave and local orbitals basis together with the Perdew-Burke-Ernzerhof parameterization of the generalized gradient approximation.[10] The plane wave cutoff parameter RMTKmax was set to 7 and the Brillouin zone (BZ) was sampled by 2000 k-points. Spin orbit coupling (SOC) was included.

**Results and discussion**

Figures 1 (a-d) show the *dc* magnetic susceptibility, the X-ray powder diffraction pattern, and the heat capacity data, respectively, for NaBi. NaBi has a tetragonal crystal structure with centrosymmetric space group P4/mmm; the diffraction pattern in panel (b) is well matched to the reported diffraction standard (ICSD No. 58816). The sharpness of peaks depicts the good crystallinity of the sample. A small peak at ~ 27.3 degrees indicates the presence of a small proportion of Bi present due to the slight decomposition of NaBi during sample grinding and data collection.

The zero-field-cooling volume susceptibility $\chi_V$ measured in a field of 5 Oe, is shown in panel (a). The experimental data, which show a full superconducting signal, have not been corrected for the demagnetization factor N ($-4\pi\chi_V = \frac{1}{1-N}$) and thus show a diamagnetism larger than 1. A sharp superconducting transition with $\Delta T_c = 0.15$ K is observed below $T_{c\ onset} = 2.15$ K. In the normal state, the magnetic susceptibility is not near zero (as is expected for weak paramagnets or diamagnets), likely caused by presence of a very small particle (a few ppm of the sample weight) of the Fe tube used for synthesis. Panel (c) presents temperature dependence of the heat capacity ($C_p/T$) in zero and 500 Oe magnetic field between 1.8 K and 2.5 K. A large heat capacity anomaly with the midpoint of superconducting transition at $T_c = 2.15$ K is evidence for the bulk character of the superconductivity in NaBi. The absence of the $C_p/T$ jump in the applied magnetic field indicates that H = 500 Oe exceeds the upper critical field for NaBi. Panel (d) shows $C_p/T$ versus $T^2$ measured in 500 Oe field. The data are well fitted by the expected formula (see figure), with the fit result represented by a red solid line. The electronic specific heat coefficient (Sommerfeld parameter) $\gamma=3.4(1)$ mJmol$^{-1}$K$^{-2}$, and phonon specific-heat coefficient $\beta = 1.44(2)$ mJmol$^{-1}$K$^{-4}$ are obtained. In the simple Debye model, the Debye temperature is given by $\Theta_D = \left(\frac{12\pi^4}{5\beta}nR\right)^{1/3}$ and using $\beta = 1.44(2)$ mJmol$^{-1}$K$^{-4}$, n = 2 and R= 8.314 J mol$^{-1}$ K$^{-1}$, the estimated Debye temperature is 140 K, comparable with $\Theta_D$ = 158 K and 119 K reported for elemental Na and Bi, respectively [11].

Using the specific heat jump shown in panel (c) $\Delta C_p/T = 2.6$ mJ mol$^{-1}$ K$^{-2}$ and taking the Sommerfeld parameter ($\gamma$), the value of $\Delta C_p/\gamma T$ can be calculated, yielding 0.78. This is about half the expected 1.43 value derived by BCS theory, and clearly indicates that NaBi is a true bulk superconductor. We attribute the discrepancy between the theory and the experimental observation to partial decomposition of the highly air sensitive material during mounting in the measurement system.

The McMillan formula for the superconducting transition temperature includes the electron-phonon coupling constant ($\lambda_{ep}$), the Debye temperature ($\Theta_D$) and the Coulomb repulsion constant ($\mu^*$):

$$T_c = \frac{\Theta_D}{1.45} \exp\left(-\frac{1.04(1+\lambda_{ep})}{\lambda_{ep} - \mu^*(1+0.62\lambda_{ep})}\right) \text{ [12]}.$$

This equation can be inverted to give $\lambda_{ep}$:

$$\lambda_{ep} = -\frac{1.04 + \mu^* \ln\left(\frac{\Theta_D}{1.45 T_c}\right)}{(1 - 0.62\mu^*)\ln\left(\frac{\Theta_D}{1.45 T_c}\right) - 1.04}$$

From the value $\Theta_D$ = 140 K, taking $T_c$ = 2.15 K and a typical $\mu^*$ = 0.13, the electron-phonon coupling constant we estimate to be $\lambda_{ep}$ = 0.62, which implies that NaBi is moderately coupled superconductor.

Having $\gamma$ and $\lambda_{ep}$ the density of states at the Fermi energy can be estimated using the relation:

$$N(E_F) = \frac{6\gamma}{\pi^2 k_B^2} \frac{1}{1+\lambda_{ep}}$$

We find $N(E_F)$ = 0.88 states eV$^{-1}$ f.u.$^{-1}$, consistent with the electronic structure calculations (see below).

The superconducting transition of NaBi was observed in electronic transport measurements as well as in magnetization measurements. Because NaBi decomposes very quickly in air, electrical contacts were made to samples while in the Ar-filled glove box. Fig. 2(a) shows the temperature dependence of the resistance of a representative sample. The residual resistivity ratio RRR ((resistance at 300 K)/(resistance at 4.2 K)) is 50, indicating a highly metallic behavior and the good quality of the NaBi sample. Fig. 2(b) shows the details of the resistance measurements near Tc; a very sharp superconducting transition appears below an onset temperature of 2.25 K. The suppression of the superconducting transition in low magnetic fields is seen in this data. The Meissner effect was observed in torque magnetometry experiments, as demonstrated in Fig. 3.

The initial magnetization curves in Fig. 3(a) show clear diamagnetic signals as the negative magnetic moment emerges below 1.9 K, and the diamagnetic signal becomes larger as the sample cools. The experimentally observed behavior is consistent with NaBi being a type II superconductor; $H_{c1}$ at 0.4 K is ~ 30 (±10) Oe. In addition, we observed a flux pinning effect in NaBi, as shown in the hysteresis loops in Fig. 3(b). To acquire the upper critical field values, we measured the magnetoresistance at different temperatures (Fig. 4(b)) of the sample whose overall resistivity behavior is shown in Fig. 2(a). At each T, $H_{C2}(T)$ was determined from the midpoints of the transitions. These are plotted in Fig. 4(a). The conventional Werthamer-Helfand-Hohenberg (WHH) theory is used to fit the $H_{C2}$ vs. T data. [13] The WHH model describes the upper critical field of a dirty type-II superconductor by

$$\ln\frac{1}{t} = \psi(\frac{1}{2} + frach2t) - \psi(\frac{1}{2})$$

where $t = \frac{T}{T_c}$, $\psi$ is the digamma function and

$$\frac{\pi^2 h}{4} = \frac{H_{c2}(T)}{(-dH_{c2}/dt)_{t=1}}.$$

From the fitting curve in Fig. 4(a), the upper critical field $H_{C2}(0)$ is determined to be $H_{C2}(0)$ = 250 Oe. Then we can calculate the coherence length with the Ginzburg-Landau formula $\xi_{Gl}(0) = \{\phi_0 /[H_{c2}(0)]\}^{1/2}$, where $\phi_0$ is the flux quanta $h/2e$; we find $\xi_{Gl}(0) = 115$ nm.

The calculated electronic band structure of NaBi is shown in Fig 5. NaBi is calculated to be good three-dimensional metal; many bands with large dispersions cross the Fermi level. The bands at the Fermi level are all derived from Bi p-orbitals. Furthermore, the SOC interaction is strong; the SOC creates large avoided crossings and effectively gaps out the Bismuth $p_z$ orbitals at the Fermi Energy ($E_F$). (The circle size in the Figure is proportional to the Bi $p_z$ orbital contribution.) The conduction at $E_F$ therefore mainly arises from the Bismuth $p_x$ and $p_y$ orbitals, with the contribution of the $p_z$ orbitals gapped out due to large SOC. In spite of the gapping of the Bi $p_z$ orbitals, the dispersion of the $p_x$ and $p_y$ orbitals is significant along z, and therefore the electronic structure can be considered as three-dimensional. Finally, the calculations yield an expected density of electronic states at the Fermi Energy ($N(E_F)$) of 0.65 states/eV/cell, which can be considered in good agreement with the experimental value of 0.88 states/eV/cell due to the fact that a small shift in the Fermi Energy will result in a significant change in the calculated $N(E_F)$.

**Conclusion**

We report methods for synthesis of the NaBi superconductor and for handling the highly air sensitive material to obtain measurements that characterize its superconducting properties. It is an excellent metal with a sharp superconducting transition and a very low upper critical field. We find its superconducting behavior at this level of analysis to be BCS-like. Electronic structure calculations indicate that spin orbit coupling has a very significant effect on the electronic states near the Fermi Energy, and that states derived from Bi orbitals dominate at $E_F$. Further characterization of this and similar alkali-Bi superconductors in the context of considering the effects of strong spin orbit coupling on superconductivity and topological physics may be of interest.

**Acknowledgements**

This research was supported by the ARO MURI on topological insulators, grant W911NF-12-1-0461 and the MRSEC program at the Princeton Center for Complex Materials, grant NSF-DMR-0819860.

**Figure Captions**

FIG. 1 (Color online) (a) The *dc* susceptibility vs. temperature for NaBi at its superconducting transition measured in an applied field of 5 Oe; (b) Room temperature powder X-ray diffraction pattern for NaBi. The red marks indicate the peak positions and intensities from ICSD No. 58816. The red arrow shows a peak from elemental Bi due to partial sample decomposition during the measurement. (c) The $C_p/T$ vs. T plot characterizing the superconducting transition. Orange solid circles are taken with no applied magnetic field, and open violet circles are taken under an applied field of 500 Oe. (d) $C_p/T$ vs. $T^2$ in the presence of the 500 Oe applied field. The open violet circles are the data points and the solid red line is the $C_p/T = \gamma + \beta T^2$ fit.

FIG. 2 (Color online) (a) The temperature dependent resistivity for NaBi over a wide temperature range. The inset (b) shows the detail of the superconducting transition in the resistivity in zero field and in various applied fields.

FIG. 3 (Color online) (a) Torque magnetometry measurements of the field dependent magnetization of NaBi in its superconducting state at different temperatures. (b) Flux pinning phenomena result in hysteresis in the torque measurements.

FIG. 4 (Color online) (a) The temperature dependence of the upper critical field, $H_{C2}(T)$, for superconducting NaBi. Blue dots show the midpoints of the superconducting transitions determined by the magnetoresistance measurements. The magenta line is the fitting curve determined by the WHH formula. It gives $H_{C2}(0) = 250$ Oe. (b) The magnetoresistance curves used to determine $H_{C2}(T)$ in the main panel.

FIG. 5 (Color online) The calculated wavevector dependent electronic band structure for NaBi, including spin-orbit coupling. The size of the circles is proportional to the Bi $p_z$ orbital contribution to the electronic structure. The inset shows the plot of the density of states with energy. The Fermi Energy is $E_F$.

**Figures**

Figure 1

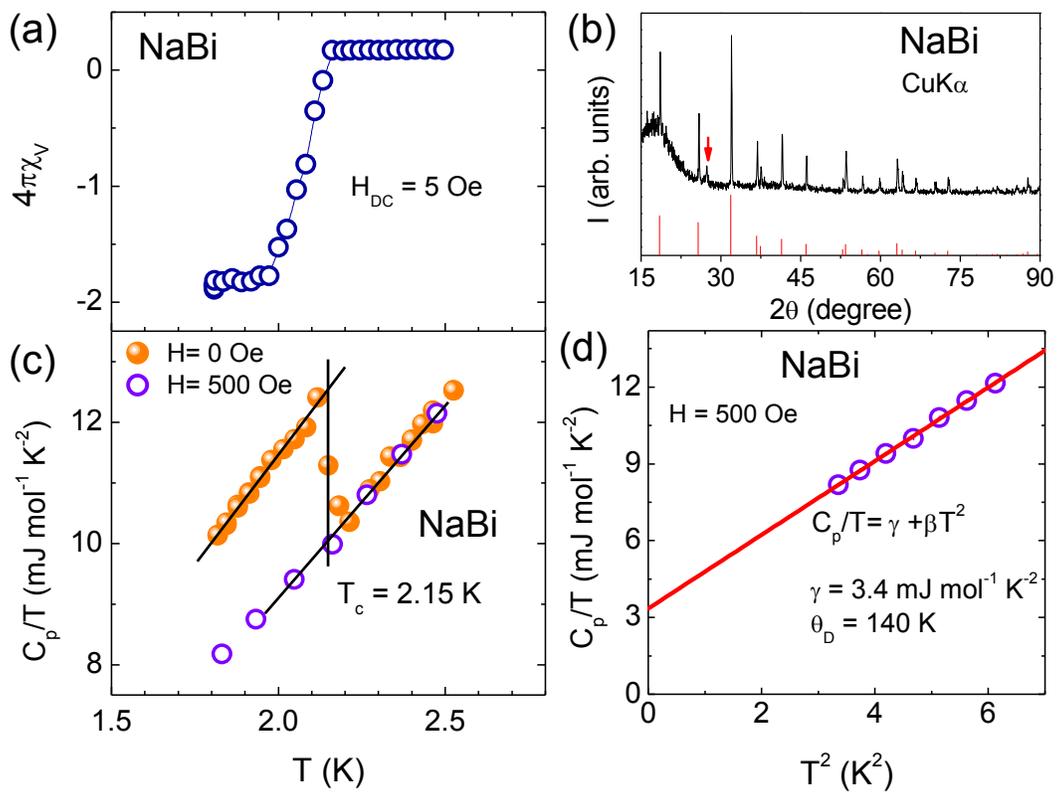

Figure 2

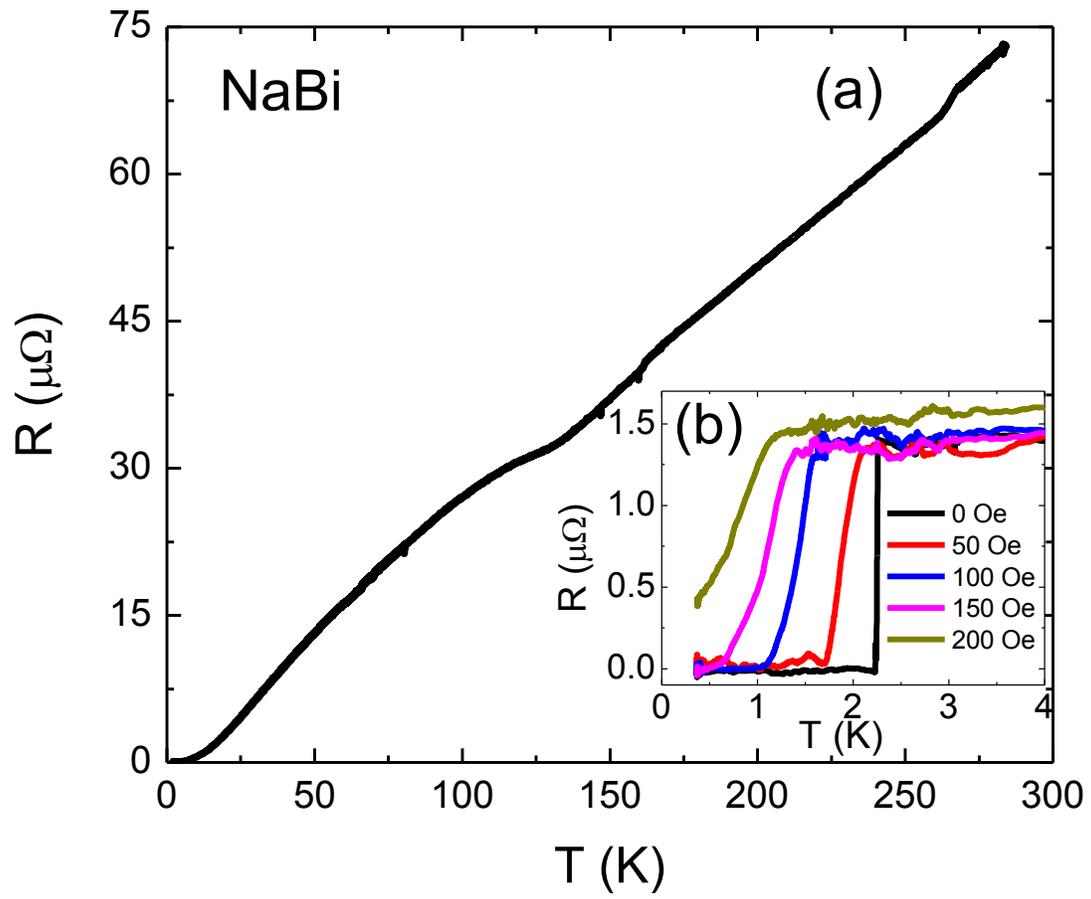

Figure 3

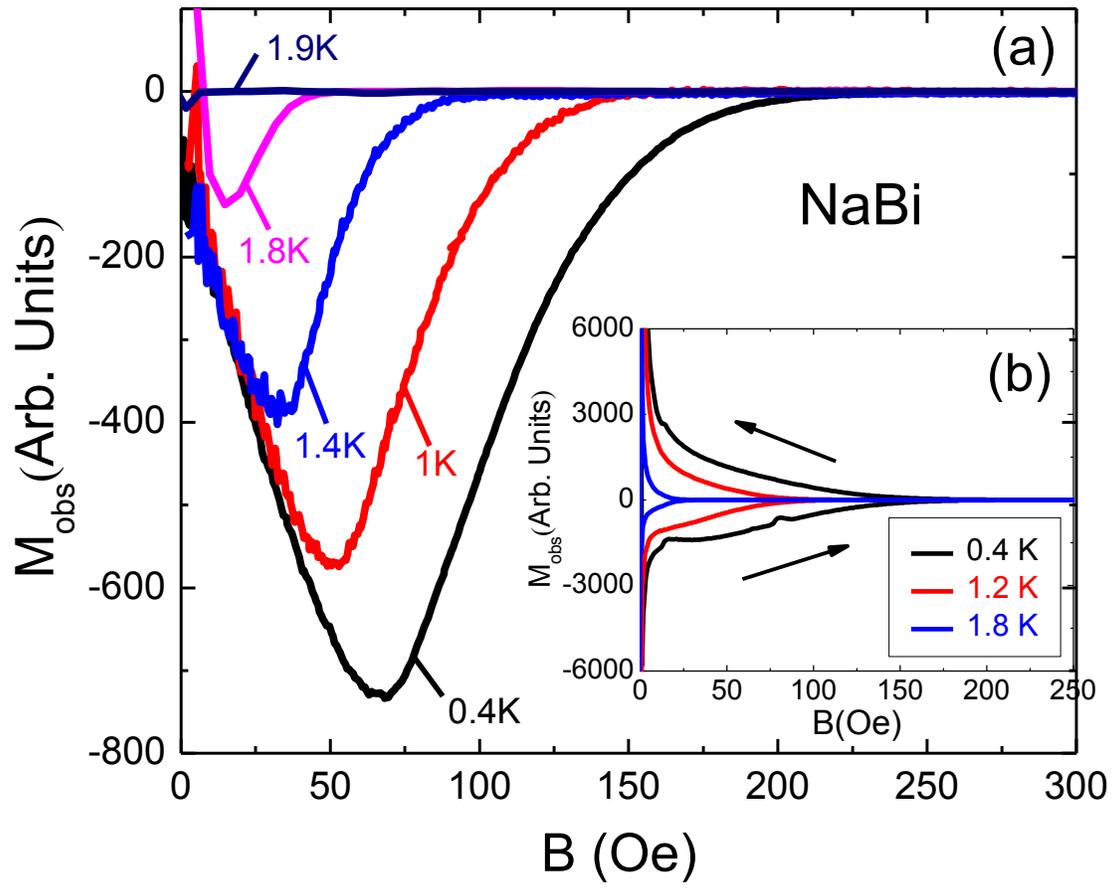

Figure 4

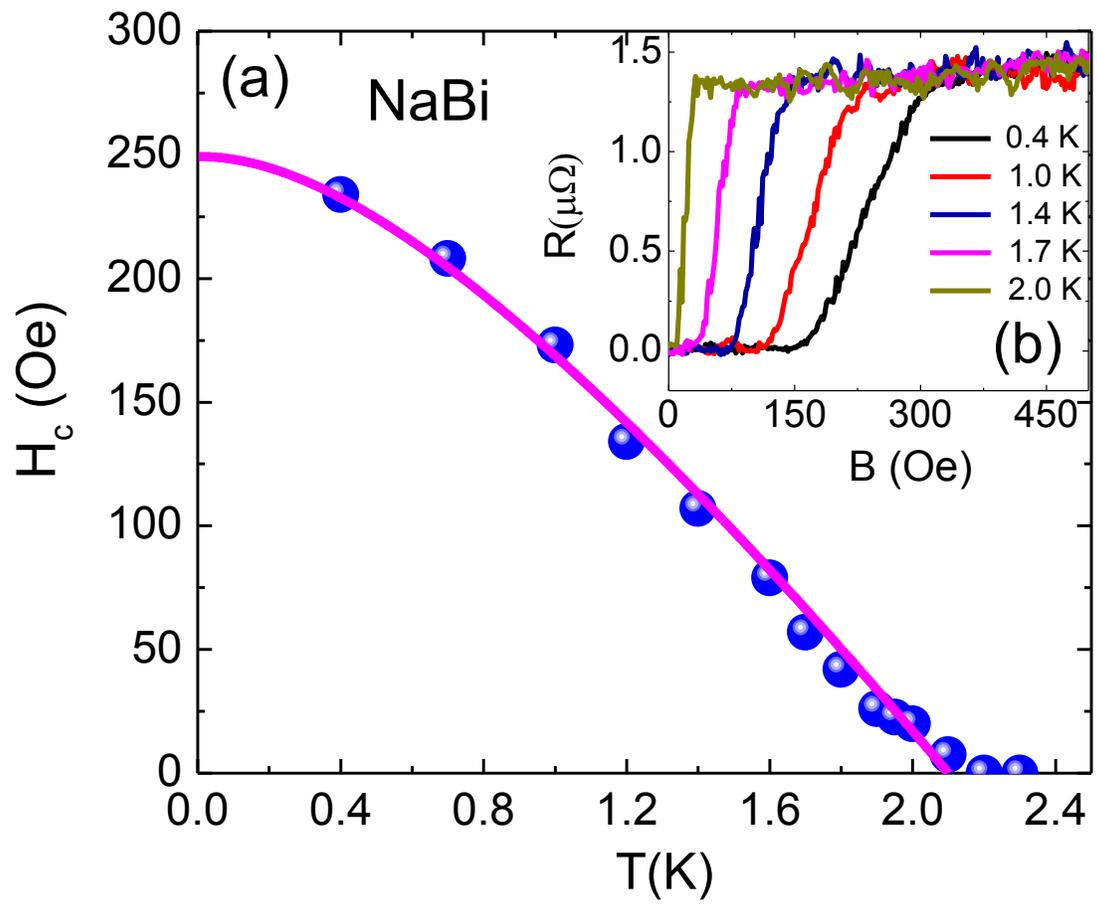

Figure 5

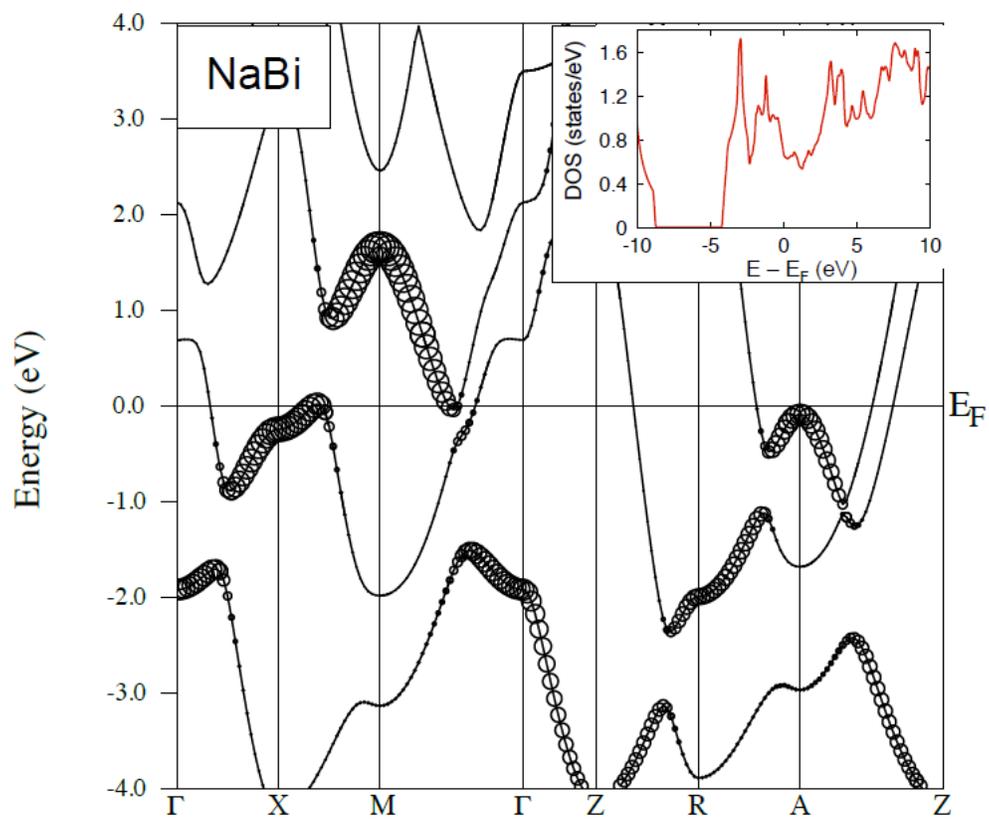